# Experimental control of quantum-mechanical entanglement in an attosecond pump-probe experiment

Lisa-Marie Koll, Laura Maikowski, Lorenz Drescher[1], Tobias Witting and Marc J.J. Vrakking*

Max-Born-Institut, Max-Born-Str. 2A, 12489 Berlin, Germany

*Corresponding author. Email: marc.vrakking@mbi-berlin.de

**Abstract:** Entanglement is one of the most intriguing aspects of quantum mechanics and lies at the heart of the ongoing Second Quantum Revolution, where it is a resource that is used in quantum key distribution, quantum computing and quantum teleportation. We report experiments demonstrating the crucial role that entanglement plays in pump-probe experiments involving ionization, which are a hallmark of the novel research field of attosecond science. We demonstrate that the degree of entanglement in a bipartite ion + photoelectron system, and, as a consequence, the degree of vibrational coherence in the ion, can be controlled by tailoring the spectral properties of the attosecond extreme ultra-violet laser pulses that are used to create them.

---

[1] Current address: Department of Chemistry, University of California, Berkeley, California 94720, USA

Within the last two decades, the field of ultrafast laser spectroscopy has been enriched by the emergence of attosecond science based on the process of high-harmonic generation (HHG) [1]. In HHG, an atomic or molecular gas is exposed to an intense, typically infrared laser, and harmonics of the driving laser frequency are produced in a three-step process [2, 3], where ionization by the intense laser field is followed by electron acceleration and electron-ion recombination accompanied by the emission of energetic photons at extreme ultra-violet (XUV) or soft X-ray wavelengths. These photons are ionizing radiation for any medium (solid, liquid or gaseous) that is placed in its path. Thus the production of attosecond pulses by HHG has sparked the emergence of a new research field, attosecond science, where studies of photoionization and of photoionization-induced dynamics play a prominent role.

Up to now, the possible role that ion-photoelectron entanglement plays in experimental attosecond science has not received prominent attention. Nevertheless, several papers have been published that have alerted to this possibility. For example, the first application of attosecond transient absorption spectroscopy (ATAS) explored coherent hole dynamics in $Kr^+$ cations formed by infrared multi-photon ionization [4], and it was pointed out that the electronic coherence in the cation was limited by entanglement with the accompanying photoelectron [5, 6]. Extending the argument to photoionization of polyatomic molecules by attosecond pulses, the limits that entanglement imposes on the observability of attosecond charge migration was discussed in [7]. Several papers have moreover theoretically discussed the role of entanglement as a limiting factor to the vibrational coherence that is observable when $H_2$ molecules are ionized by an attosecond pulse [8, 9]. A review of the possible role of quantum mechanical entanglement in atomic and molecular (AMO) physics is given in [10].

A multi-component quantum system is entangled, when it is impossible to represent its state in terms of a single product of wave functions describing the individual parts. For example, in the case of photoionization, the wave function of an entangled ion + photoelectron pair is written as a sum of such products, i.e., $\Psi \sim \sum_{\alpha,\beta} \psi_{ion,\alpha} \psi_{photoelectron,\beta}$, where $\alpha$ and $\beta$ represent the full set of (discrete and continuous) quantum numbers needed to characterize the ionic wave function $\psi_{ion}$ and the photoelectron wave function $\psi_{photoelectron}$. Entanglement naturally arises in molecular photoionization, since the electronic and ro-vibrational states of the cation, and the energy and angular momentum of the photoelectron occur in different combinations and remain uncertain until they are determined by a measurement. In an entangled system, knowledge obtained about one part of the system by a measurement (or even just the possibility to perform such a measurement), has measurable consequences for what can be observed in the other part.

The entanglement that arises in photoionization limits the coherence properties of the ion and photoelectron that are produced. Here, we demonstrate this by considering the example of H$_2$. As previously experimentally demonstrated, photoionization of H$_2$ by an attosecond pulse can lead to the formation of a coherent superposition of vibrational levels in the $1s\sigma_g$ state of H$_2^+$ [11], corresponding to the formation of a vibrational wave packet that oscillates back and forth between the inner and outer turning point of the $1s\sigma_g$ potential energy curve. The fact that this cationic state is in a coherent superposition can be confirmed by a pump-probe experiment, where a short near-infrared (NIR) probe pulse dissociates the molecule at a variable time delay with respect to the photoionization event, promoting the wave packet from the bound $1s\sigma_g$ to the dissociative $2p\sigma_u$ electronic state, and producing an easily measurable H$^+$ fragment ion.

Following a recent proposal [9], we experimentally demonstrate control of the degree of entanglement at the expense of the vibrational coherence of the H$_2^+$ ion, by performing pump-probe experiments where H$_2$ was ionized by a pair of attosecond pulses with a variable relative delay $\tau_{\text{XUV-XUV}}$. In order to motivate the experiment, it is useful to consider just two of the cationic vibrational levels $H_2^+(v)$ and $H_2^+(v')$. Interference between the two attosecond pulses produces a fringe pattern in the XUV spectrum, with a fringe spacing $\Delta\omega$ that is inversely proportional to the two-pulse delay according to $\Delta\omega = \frac{2\pi}{\tau_{\text{XUV-XUV}}}$ (a.u.). When $\tau_{\text{XUV-XUV}} = \frac{2n\pi}{\Delta E_{v,v'}}$, where $\Delta E_{v,v'}$ is the energy separation between levels $H_2^+(v)$ and $H_2^+(v')$, and $n$ is an integer, the photoelectron spectra accompanying the formation of $H_2^+(v)$ and $H_2^+(v')$ are shifted with respect to each other by $n$ fringes. If the bandwidth of the ionizing light source significantly exceeds $\Delta E_{v,v'}$, which is naturally the case for an attosecond pulse pair, the photoelectron spectra largely overlap. Then, we can approximate the ion + photoelectron wave function as $\Psi = \psi_{\text{ion},v}\psi_{\text{photoelectron},v} + \psi_{\text{ion},v'}\psi_{\text{photoelectron},v'} \approx (\psi_{\text{ion},v} + \psi_{\text{ion},v'})\psi_{\text{photoelectron}}$, and the coherent superposition of cationic vibrational states $H_2^+(v)$ and $H_2^+(v')$ is observable in a pump-probe experiment, where pathways towards the final state involving $\psi_{\text{ion},v}$ and $\psi_{\text{ion},v'}$ interfere constructively or destructively as a function of pump-probe delay. Conversely, when $\tau_{\text{XUV-XUV}} = \frac{(2n+1)\pi}{\Delta E_{v,v'}}$, then the photoelectron spectrum accompanying the formation of $H_2^+(v)$ is shifted by a half-integer number $(2n+1)/2$ fringes with respect to the photoelectron spectrum accompanying $H_2^+(v')$, implying that a possible measurement of the photoelectron kinetic energy permits to distinguish the formation of $H_2^+(v)$ and $H_2^+(v')$ with a high chance of success. In this case, no further simplification of the entangled wave function $\Psi = \psi_{\text{ion},v}\psi_{\text{photoelectron},v} + \psi_{\text{ion},v'}\psi_{\text{photoelectron},v'}$ is possible and the ion + photoelectron entanglement is expected to limit the detection of a coherent superposition of vibrational levels $H_2^+(v)$ and $H_2^+(v')$. The criterion that $\tau_{\text{XUV-XUV}} = \frac{(2n+1)\pi}{\Delta E_{v,v'}}$ leads to entanglement can also be derived by determining the conditions under which the off-diagonal elements of the reduced ionic density matrix, which are a measure of the vibrational coherence, vanish. As derived in the Supplemental Material, these matrix-elements are given by

$$\rho_{vv'} \sim \int d\varepsilon\, F_{0,\text{XUV}}(\varepsilon + E_v)\, F_{0,\text{XUV}}^*(\varepsilon + E_{v'})$$
$$\cos\left[\frac{\Delta E_{v,v'}\tau_{\text{XUV-XUV}}}{2}\right] \cos\left[\left(\frac{(E_v + E_{v'})}{2} + \varepsilon\right)\tau_{\text{XUV-XUV}}\right]$$

(1)

where $F_{0,\text{XUV}}$ is the spectral amplitude of the individual XUV pulses and $\varepsilon$ the photoelectron energy. Hence, $\rho_{vv'}$ is zero whenever $\frac{\Delta E_{v,v'}\tau_{\text{XUV-XUV}}}{2} = \frac{(2n+1)\pi}{2}$.

To prepare a pair of attosecond pulses with variable delay, two collinearly propagating sub-5 fs NIR pulses were produced using an ultra-stable Mach-Zehnder interferometer with a motorized delay stage in one of the arms (see Figure 1). The interferometer employed a combination of passive and active stabilization, and was constructed using optics compatible with few-cycle NIR pulses. Passive stabilization was achieved by constructing the interferometer on a vibrationally decoupled breadboard, and by mounting the retro-reflecting mirror pairs on thick stainless steel plates, thus shifting the resonance frequency below prominent laboratory noise sources. This approach led to a short term delay stability well below 10 as (1 as = $10^{-18}$ s). Active stabilization was implemented using a frequency-stable, 473 nm cw laser that co-propagated in both interferometer arms, with a 30 arcmin wedge inserted into one of the two arms to induce tilt fringes in the plane of detection. Path length variations were measured by Fast Fourier Transform (FFT) processing of fringe patterns recorded by a fast USB3 CMOS camera, using the Takeda algorithm [12], and were used to derive an error signal that was fed into a proportional–integral–derivative (PID) controller used to drive the delay stage. Thus, two-pulse delay fluctuations could be kept below 10 as r.m.s. for multiple days while scanning $\tau_{\text{XUV-XUV}}$ over a range >100 fs.

The NIR pulse pair was used to generate high harmonics in Ar, within an attosecond pump-probe setup that has been previously described [13, 14] (see Supplemental Material for a concise summary). Two-pulse delays $\tau_{\text{XUV-XUV}}$ between 11 and 102 fs (in steps of 3 fs) were employed, with the lower limit chosen to avoid optical interferences between the two NIR pulses. Inspection of the HHG spectra recorded under these conditions (see Figure 1) revealed clear delay-dependent spectral modulations with a fringe spacing inversely proportional to $\tau_{\text{XUV-XUV}}$, demonstrating that the HHG by the 2nd NIR pulse was not appreciably affected by modifications of the generating medium by the 1st NIR pulse. The XUV pulses were collinearly overlapped with a co-polarized replica of the NIR pulse, and focused into a velocity map imaging (VMI) spectrometer [15], where both pulses crossed a pulsed $H_2$ jet generated within the repeller electrode [16]. $H^+$ ions resulting from two-color XUV+NIR dissociative ionization were projected onto a 2D MCP +

phosphor screen detector, and recorded using a CCD camera. The XUV and NIR polarizations were in the plane of the MCP detector, permitting use of an inverse Abel transform (see Supplemental Material) to retrieve the three-dimensional (3D) H$^+$ momentum distributions. Images were recorded as a function of $\tau_{XUV-XUV}$, and as a function of the XUV-NIR time delay $\tau_{XUV-NIR}$, which was varied between -50 and +800 fs in steps of 4 fs. For each $\tau_{XUV-XUV}$ and $\tau_{XUV-NIR}$, up to 8 pump-probe scans were performed, where data was collected for 2000 laser shots at each delay.

An example of a measured and an inverted image are shown in Figure 1. They show that the H$^+$ momentum distribution consists of a contribution at low momentum resulting from single photon XUV dissociative ionization (where the dissociation into H$^+$ + H takes place on the $1s\sigma_g$ potential energy curve) and a contribution at higher momentum resulting from 1+1' XUV+NIR two-color dissociative ionization (where the dissociation takes place on the $2p\sigma_u$ potential energy curve). The former signal was used for normalization purposes, since it is linearly proportional to both the XUV flux and the H$_2$ gas density.

Considering that the 3D H$^+$ momentum distributions retrieved by Abel inversion only depend on the absolute H$^+$ velocity $v_{H^+}$, and the angle $\theta$ between this velocity and the XUV and NIR polarization axis, the experiment led to the determination of a 4D probability distribution $P(v_{H^+}, \theta, \tau_{XUV-XUV}, \tau_{XUV-NIR})$. Since the XUV+NIR dissociative ionization is a two-photon process, this probability distribution can be written as

$$P(v_{H^+}, \theta, \tau_{XUV-XUV}, \tau_{XUV-NIR}) = \beta_0(v_{H^+}, \tau_{XUV-XUV}, \tau_{XUV-NIR}) \times$$

$$\{1 + \beta_2(v_{H^+}, \tau_{XUV-XUV}, \tau_{XUV-NIR}) P_2(cos\theta) + \beta_4(v_{H^+}, \tau_{XUV-XUV}, \tau_{XUV-NIR}) P_4(cos\theta)\}$$

where $P_2(cos\theta)$ and $P_4(cos\theta)$ are 2$^{nd}$ order and 4$^{th}$ order Legendre polynomials. Figure 2(a)-(f) show experimental results for $\beta_0(v_{H^+}, \tau_{XUV-XUV}, \tau_{XUV-NIR})$, $\beta_2(v_{H^+}, \tau_{XUV-XUV}, \tau_{XUV-NIR})$ and $\beta_4(v_{H^+}, \tau_{XUV-XUV}, \tau_{XUV-NIR})$ as a function of $v_{H^+}$ and $\tau_{XUV-NIR}$ for two different values $\tau_{XUV-XUV}$ = 29 and 45 fs. The measurements show wave packet motion with a vibrational period of almost 30 fs, and reveal the dephasing and rephasing that is commonly observed when coherent superpositions consisting of ≥3 states are excited. A profitable way to analyze these results is by performing a Fourier Transform over $\tau_{XUV-NIR}$, and plotting the Fourier Transform Power Spectrum $FTPS(v_{H^+}, \tau_{XUV-XUV}, \omega)$ where $\omega$ is the Fourier frequency. FTPSs for the data shown in Figures 2(a)-(f) are presented in Figures 2(g)-(l). They reveal that the observed dependence of $\beta_0(v_{H^+}, \tau_{XUV-XUV}, \tau_{XUV-NIR})$, $\beta_2(v_{H^+}, \tau_{XUV-XUV}, \tau_{XUV-NIR})$ and $\beta_4(v_{H^+}, \tau_{XUV-XUV}, \tau_{XUV-NIR})$ on $\tau_{XUV-NIR}$ is determined by a number of nearest-neighbor and next-nearest-neighbor quantum beats. Column 2 of Table I presents an overview of the most prominent beat frequencies that are observable in Figure 2, which can readily be assigned using available spectroscopic information (last column) [17]. Importantly, whereas the contributions of next-nearest neighbor quantum beats are of comparable strength for $\tau_{\text{XUV-XUV}}$ = 29 and 45 fs, the nearest-neighbor quantum beats are much more prominent in the former measurement, and only weakly visible in the latter.

For each $\tau_{\text{XUV-XUV}}$ between 11 and 102 fs, a scan like the one shown in Figure 2 was performed, allowing a determination of the intensity of the nearest-neighbor and next-nearest neighbor quantum beats as a function of $\tau_{\text{XUV-XUV}}$. This intensity dependence is shown in Figure 3, and represents the main result of our work. Figure 3 shows that the intensity of each two-level quantum beat in the FTPS oscillates as a function of $\tau_{\text{XUV-XUV}}$ with a period (see figure) inversely proportional to the frequency of the quantum beat. Thus, Figure 3 provides experimental confirmation that vibrational quantum beats are observable when $\tau_{\text{XUV-XUV}} = \frac{2n\pi}{\Delta E_{v,v\prime}}$ and are absent or only weakly observable when $\tau_{\text{XUV-XUV}} = \frac{(2n+1)\pi}{\Delta E_{v,v\prime}}$, with $n$ an integer. Values of $\Delta E_{v,v\prime}$ extracted from the fits shown in Figure 3 by means of a weighted averaging are shown in the 3$^{\text{rd}}$ column of Table I. As before, they show reasonable agreement with literature results.

The behavior in Figure 3 is caused by quantum mechanical entanglement of the $H_2^+$ ion with the photoelectron that accompanies its formation. Although we have focused our attention on the vibrational coherence in the $H_2^+$ ion within the pump-probe experiment, this ion remains part of an entangled system throughout the experiment. Modifying the spectral properties of the attosecond XUV pulse pair modifies the photoelectron spectra accompanying the different $H_2^+$ vibrational states, and thus affects the degree of ion + photoelectron entanglement, thereby controlling the observability of the $H_2^+$ vibrational coherences.

It should be acknowledged and emphasized that the use of a two-pulse sequence to tailor excited state properties is nothing new [18]. This approach underlies Ramsey-type Fourier Transform spectroscopy [19], has been exploited in coherent control experiments [20] and plays an essential role in applications ranging from multi-dimensional spectroscopy [21] to hyperspectral imaging [22, 23]. The unique aspect of our experiment is that the two-pulse sequence is not used to control excited state populations, but is used to control the entanglement in a bipartite system consisting of an ion and a photoelectron. Given that these states are formed by an ionization process, rather than a resonant bound-to-bound transition, the cationic vibrational state populations do not depend at all on the two-pulse delay [9], and, except for the narrow fringe patterns caused by $\tau_{XUV-XUV}$, the same is true for the photoelectron spectrum that accompanies the ionization process. What radically changes as a function of $\tau_{XUV-XUV}$ is the ion + photoelectron entanglement. This entanglement determines the observability of quantum beats between pairs of vibrational states within the pump-probe experiment. Whenever the photoelectron spectrum accompanying both vibrational states is very similar (i.e., offset by an integer number of fringes in the XUV spectrum), the experiment reveals pronounced quantum beats, whereas these quantum beats are greatly suppressed when the XUV fringe pattern produces a situation where the photoelectron spectrum accompanying both vibrational states under consideration is different.

Ion + photoelectron entanglement is a so far underappreciated aspect of pump-probe experiments involving ionization, which are ubiquitous in attosecond science. Indeed, it may be expected that ion + photoelectron entanglement is highly common within attosecond science. As an example, we mention our previous work on laboratory frame electron localization in $H_2$ [24], which was a first example of attosecond pump-probe spectroscopy to molecules, and which can be described within a very similar conceptual framework as the work presented here. Electron localization in the dissociation of $H_2^+$ into $H^+ + H$ requires that the parity of the ionic wave function is broken, i.e., that the ion is produced in a coherent superposition of gerade and ungerade states. However, photoionization produces the $H_2^+$ ion as part of a bipartite entangled state, where formation of gerade ionic states is accompanied by the emission of an "ungerade" photoelectron with odd orbital angular momentum, and formation of ungerade ionic states is accompanied by the emission of a "gerade" photoelectron with even angular momentum. The ionic coherence that is needed to observe laboratory frame electron localization is absent in this entangled state and can be induced within a pump-probe experiment when interactions of the ion or the photoelectron with the probe laser reduce the degree of entanglement. Indeed, both mechanisms were observed in [24].

Our work represents both a warning and an opportunity. It is a warning, because it demonstrates how ion + photoelectron entanglement may prevent the observation of coherent dynamics in ions or photoelectrons, which is the objective of many attosecond experiments. It is also an opportunity, because it shows that tailoring the properties of the attosecond laser pulses can be used to emphasize the role of selected two-level quantum beats. This may be of particular interest in current efforts targeting the observation of attosecond to few-femtosecond charge migration and charge-directed reactivity [25, 26], where initial experimental results have suggested the key role of particular two-level quantum beats [27]. Finally, our work draws attention to the links that may be established between ultrafast laser spectroscopy and the field of quantum information [10], where the experimental tools of attosecond science may create hitherto unsuspected opportunities.

**Figures**

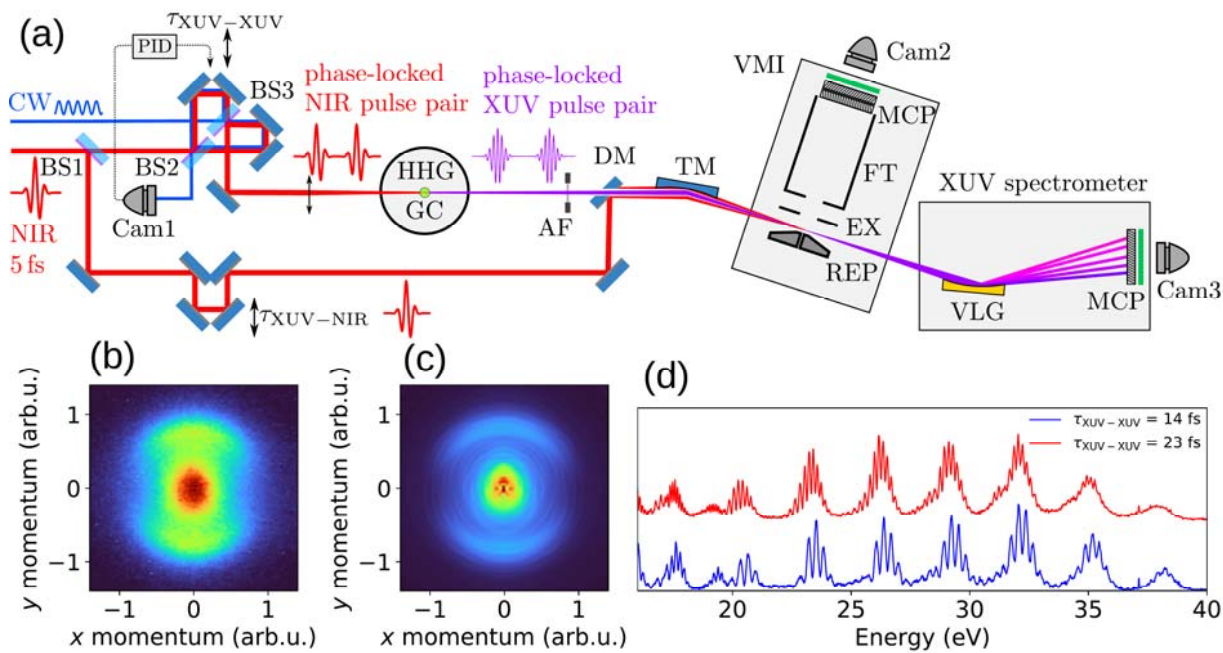

**Fig. 1. Setup for measuring the influence of quantum mechanical entanglement on vibrational wave packet dynamics in $H_2^+$.** a) Neutral $H_2$ molecules injected into the apparatus through a nozzle incorporated in the repeller (REP) of the velocity map imaging (VMI) spectrometer, are dissociatively ionized using a two-color XUV+NIR laser field, where the XUV field consists of a two-pulse pair with a relatively delay $\tau_{XUV-XUV}$, which is followed by the NIR pulse after a delay $\tau_{XUV-NIR}$. The XUV pulse-pair is generated by passing ultrashort NIR pulses through a passively and actively stabilized Mach-Zehnder interferometer prior to their use within a high-harmonic generation (HHG) gas cell (GC). The XUV and NIR pulses are collinearly combined using a drilled mirror (DM) and focused using a toroidal mirror (TM). The two-pulse XUV spectrum is monitored using an XUV spectrometer. Abbreviations: BS1-BS3 beamsplitters; PID proportional–integral–derivative feedback for stabilization of $\tau_{XUV-XUV}$; Cam1-3 Cameras; AF Aluminum filter; EX Extractor electrode; FT Flight tube; MCP Microchannel Plate; VLG Variable Line-space Grating; b) and c) typical example of a raw VMI image and a slice through the 3D momentum distribution obtained via Abel inversion; d) XUV spectrum recorded for two delays $\tau_{XUV-XUV}$, revealing fringes in the XUV spectrum resulting from the interference between the two XUV pulses.

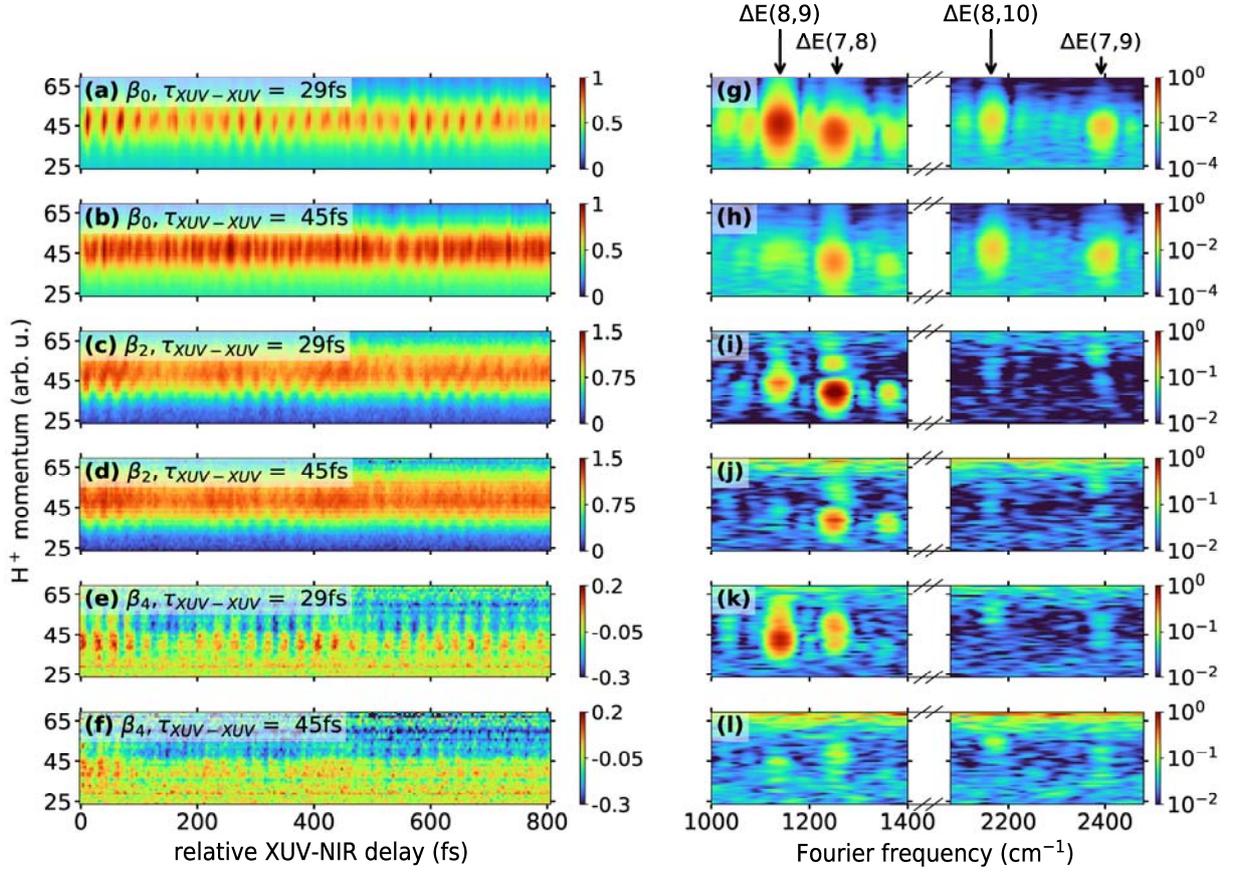

Fig. 2. Experimentally observed wave packet dynamics following ionization of $H_2$ by a pair of XUV pulses separated by $\tau_{\text{XUV}-\text{XUV}} = 29$ and 45 fs: a-f) $\beta_0(v_{H^+}, \tau_{\text{XUV}-\text{XUV}}, \tau_{\text{XUV}-\text{NIR}})$, $\beta_2(v_{H^+}, \tau_{\text{XUV}-\text{XUV}}, \tau_{\text{XUV}-\text{NIR}})$ and $\beta_4(v_{H^+}, \tau_{\text{XUV}-\text{XUV}}, \tau_{\text{XUV}-\text{NIR}})$ obtained by Abel inversion of the experimental VMI images. The vertical axis is given in pixel units of the CCD camera used in the VMI spectrometer, and represents the $H^+$ velocity in arbitrary units; g-l) Fourier Transform Power Spectra $FTPS(v_{H^+}, \tau_{\text{XUV}-\text{XUV}}, \omega)$ of $\beta_0(v_{H^+}, \tau_{\text{XUV}-\text{XUV}}, \tau_{\text{XUV}-\text{NIR}})$, $\beta_2(v_{H^+}, \tau_{\text{XUV}-\text{XUV}}, \tau_{\text{XUV}-\text{NIR}})$ and $\beta_4(v_{H^+}, \tau_{\text{XUV}-\text{XUV}}, \tau_{\text{XUV}-\text{NIR}})$, obtained by a Fast Fourier Transform. The vertical axis coincides with the vertical axis in a-f), whereas the horizontal axis gives the Fourier frequency in cm$^{-1}$, permitting a straightforward assignment to two-level quantum beats and a comparison with available literature values (see Table I). Importantly, the intensity of the nearest-neighbor quantum beats in the FTPS are significantly different for the two different values of $\tau_{\text{XUV}-\text{XUV}}$.

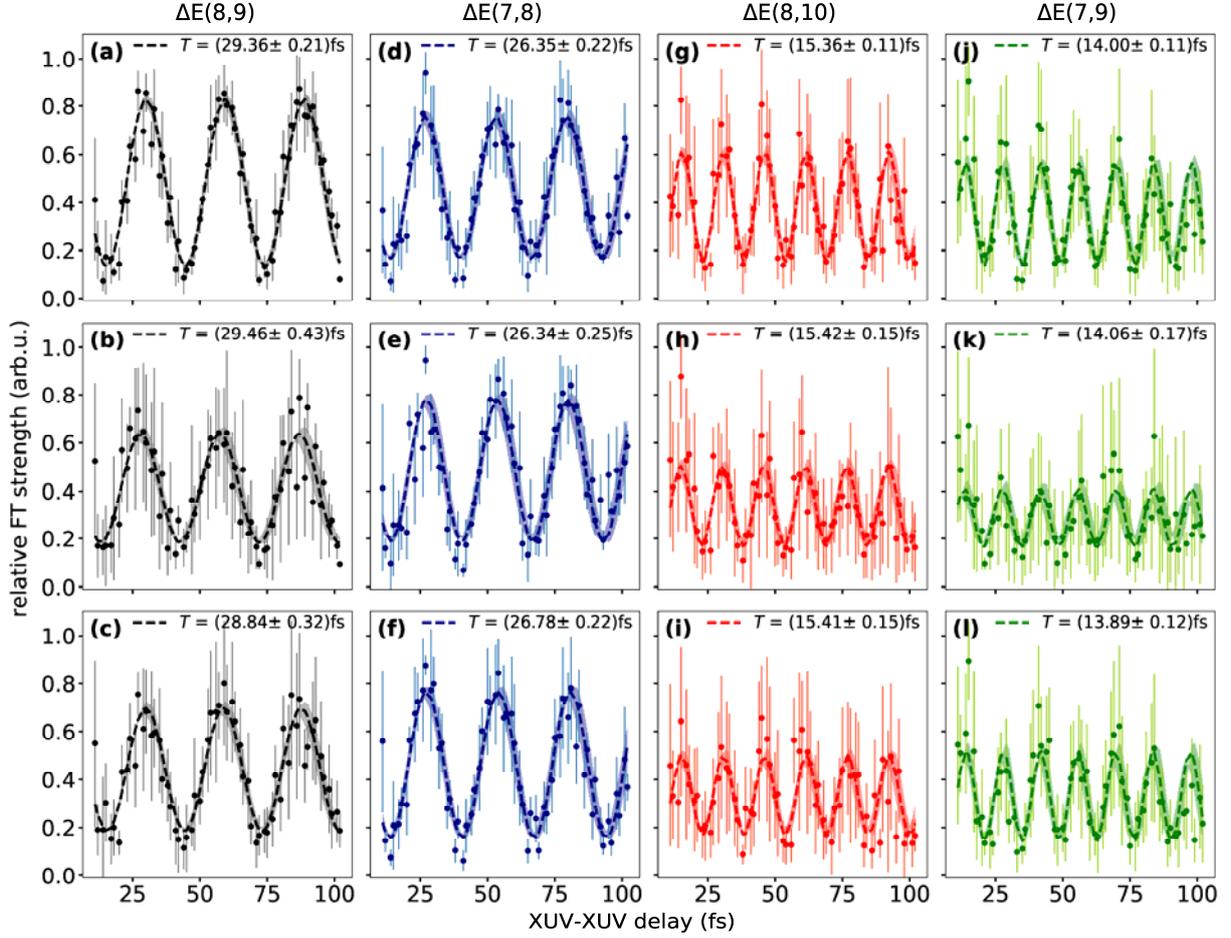

**Fig. 3. Intensities of peaks in the Fourier Transform Power Spectrum $FTPS(v_{H^+}, \tau_{XUV-XUV}, \omega)$ as a function of two-pulse delay $\tau_{XUV-XUV}$, for 4 selected values of the Fourier frequency:** a)-c) $FTPS_{\beta_0}(v_{H^+}, \tau_{XUV-XUV}, \omega)$, $FTPS_{\beta_2}(v_{H^+}, \tau_{XUV-XUV}, \omega)$ and $FTPS_{\beta_4}(v_{H^+}, \tau_{XUV-XUV}, \omega)$ with $\omega$ corresponding to the energy difference between $v^+=8$ and 9; d)-f) the same, with $\omega$ corresponding to the energy difference between $v^+=7$ and 8; g)-i), the same, with $\omega$ corresponding to the energy difference between $v^+=8$ and 10; j)-l) the same, with $\omega$ corresponding to the energy difference between $v^+=7$ and 9. The oscillations in the FTPS (experimental data points given by symbols) are fitted to sinusoidal curves (dashed lines, with shaded area according to the standard deviation of the fit parameters), leading to a determination of the oscillation periods shown at the top of each figure. The oscillation periods obtained for $FTPS_{\beta_0}(v_{H^+}, \tau_{XUV-XUV}, \omega)$, $FTPS_{\beta_2}(v_{H^+}, \tau_{XUV-XUV}, \omega)$ and $FTPS_{\beta_4}(v_{H^+}, \tau_{XUV-XUV}, \omega)$ are combined to obtain value that are compared to available spectroscopic data in column 3 of Table I.

| Fourier Peak (Figure 1) | Fourier Frequency obtained by FFT of pump-probe delay scans (cm$^{-1}$) (see Figure 2) | Fourier Frequency obtained by FFT of the $\tau_{XUV-XUV}$ delay dependence of the FTPS (cm$^{-1}$) (see Figure 3) | Assignment (v,v′) | Literature value (cm$^{-1}$) [a] |
|---|---|---|---|---|
| A | 1138 ± 7 | 1141 ± 7 | (8,9) | 1130.1 |
| B | 1252 ± 7 | 1259 ± 6 | (7,8) | 1262.5 |
| C | 2168 ± 16 | 2167 ± 11 | (8,10) | 2127.8 |
| D | 2396 ± 17 | 2387 ± 13 | (7,9) | 2392.6 |

**Table 1. Comparison of experimental Fourier frequencies with literature values.** Column 2 was obtained by performing a statistical analysis for the measured $FTPS(v_{H^+}, \tau_{XUV-XUV}, \omega)$ after integrating over $v_{H^+}$, using the data for $\beta_0$ only (see Fig. 2). Column 3 was obtained by fitting the peak intensities in the FTPS as a function of $\tau_{XUV-XUV}$ to a sinusoidal curve (see Fig. 3).

Supplemental Material for:

# Experimental control of quantum-mechanical entanglement in an attosecond pump-probe experiment

Lisa-Marie Koll, Laura Maikowski, Lorenz Drescher, Tobias Witting and Marc J.J. Vrakking

Max-Born-Institut, Max-Born-Str. 2A, 12489 Berlin, Germany

**Attosecond Setup**

The laser system used in the experiments was a Ti:sapphire laser system (Aurora, Amplitude Technologies) operating at 1 kHz repetition rate and providing pulse energies up to 20mJ, with a pulse duration of typically 28 fs (as characterized by a spectral interferometry for direct electric field reconstruction (SPIDER) measurement [1]).

In order to generate near-single cycle pulses, up to 2.0 mJ were sent into a differentially pumped hollow fiber compression system [2]. The fused silica hollow waveguide had an inner diameter of 340 μm and was 1m long. Neon was used as the non-linear medium, with a 2 bar pressure at the exit side. The output pulses had an energy of up to 1.4 mJ and were compressed by double angle technology chirped mirrors (PC70, Ultrafast Innovations GmbH) [3, 4]. The third order dispersion (TOD) was compensated by transmitting the beam through a 2 mm long z-cut Potassium Dideuterium Phosphate (KDP) crystal. The dispersion was fine-tuned by a pair of fused silica wedges (OA925, Spectra-Physics Vienna). The pulse compression was monitored and optimized using the spatially encoded arrangement for direct electric field reconstruction by spectral shearing interferometry (SEA-F-SPIDER) technique [5]. In the present experiments the pulse duration was 5 fs.

The compressed pulses were sent to a Mach-Zehnder type interferometer where 20% of the pulse energy was split off (see BS1 in Fig1 of the main paper) to serve as a probe pulse, which can be delayed by a stick-slip piezo stage (SLC-1730-LC-ST, Smaract GmbH). The main part of the energy was sent into a compact dispersion-balanced Mach-Zehnder interferometer (see Fig1, main manuscript). This interferometer creates a pair of phase-locked NIR pulses. The delay between the two NIR pulses can be continuously adjusted over a range of 400 fs with a delay stability of <10 as rms. Further details of the phase-stabilization system are described in the main paper.

The phase-locked pair of NIR pulses were focused by a spherical mirror (f = 750 mm) into a 3 mm diameter gas cell filled with 40 mbar of argon gas, placed inside a vacuum chamber ('HHG' in Fig 1, main paper). The XUV beam was transmitted through a 200 nm thin foil of aluminum to block the residual NIR pulses and provide negative group velocity dispersion (GVD) for attochirp compensation [6]. The NIR pulses from the probe arm were recombined with the XUV pulse pairs using a drilled mirror. Both the NIR and XUV foci were re-imaged into the velocity map imaging (VMI) chamber with a gold-coated toroidal mirror in a 2f-2f arrangement. The XUV spectra were monitored with a flat-field XUV spectrometer [7] equipped with a Variable Line-space Grating (VLG, 001-0640, Hitachi).

In the experiments presented in this paper, H+ ions were detected in a VMI spectrometer with repeller-integrated gas jet [8].

**Abel Inversion**

In the experiment, $H^+$ ions formed at the crossing point of the $H_2$ gas jet and the XUV and NIR laser beams were accelerated towards a two-dimensional (2D) detector that consists of a dual micro-channel plate stack, followed by a phosphor screen and a camera system that records the positions of the particle impacts. Since the three-dimensional (3D) momentum distribution contains an axis of symmetry in the plane of the detector, the measurement is an Abel projection, and the 3D momentum distribution can be retrieved from the 2D projection by means of an inverse-Abel transform [9].

Angular distributions that occur in multi-photon dissociative ionization can be expressed as a superposition of Legendre polynomials, where the highest-order polynomial involved is determined by the order of the multi-photon ionization process [10]. It follows that, whereas the measurement is typically performed using a camera system that delivers signal intensities on a Cartesian grid, the Abel inversion problem itself is more favorably formulated using Legendre polynomials to describe the 2D measured and the 3D reconstructed momentum distribution. Doing so turns the inversion problem into a simple and fast matrix multiplication.

The cylindrically symmetric 3D momentum distribution that needs to be determined can be expressed by

$$P_{3D}(v_{3D}, \theta_{3D}, \varphi_{3D}) = \sum_l a_{v_{3D},l} P_l(\cos\theta_{3D}) \tag{1}$$

where $v_{3D}$ is the particle velocity and $\theta_{3D}$ and $\varphi_{3D}$ are angles describing the velocity of the particle with respect to the symmetry axis of the experiment, which generally coincides with the polarization axis of the light source. $\theta_{3D}$ takes on values between 0 and $\pi$, while $\varphi_{3D}$ varies between 0 and $2\pi$.

When $P_{3D}(v_{3D}, \theta_{3D}, \varphi_{3D})$ is expressed by equation (1), the particle velocity distribution $P(v)$ and the particle kinetic energy distribution $P(E)$ can be readily obtained, according to

$$P(v) = 4\pi v_{3D}^2 a_{v_{3D},l=0} \qquad (2)$$

and

$$P(E) = (4\pi/m) v_{3D} a_{v_{3D},l=0} \qquad (3)$$

Similarly, the β-parameters describing the angular distribution of the H$^+$ fragment ions can be readily obtained as

$$\beta_{2,v} = a_{v_{3D},l=2}/a_{v_{3D},l=0} \qquad (4)$$

and

$$\beta_{4,v} = a_{v_{3D},l=4}/a_{v_{3D},l=0} \qquad (5)$$

The projection of the H$^+$ fragment ions on the detector leads to a 2D momentum distribution that can similarly be expressed in terms of Legendre polynomials according to

$$P_{2D}(v_{2D}, \theta_{2D}) = \sum_l b_{v_{2D},l} P_l(\cos\theta_{2D}) \qquad (6)$$

where $v_{2D}$ is the measured velocity in the plane of the detector and $\theta_{2D}$ is the angle of this velocity with respect to the symmetry axis (i.e., the polarization axis of the lasers).

Abel projection of the 3D momentum distribution onto the 2D detection plane implies that the array of elements $b_{v_{2D},l}$ in Eqn. (5), which fully describe the 2D projection (and that can be organized into a vector **b**), can be obtained from the array of elements $a_{v_{3D},l}$ in Eqn. (1) (organized into a vector **a**) by means of a matrix multiplication:

$$\mathbf{b} = \mathbf{M}\,\mathbf{a} \qquad (7)$$

Following a determination of the inverse matrix **M**$^{-1}$, the array describing the 3D momentum distribution **a** can be easily and quickly obtained from the array describing the 2D projection **b** according to:

$$\mathbf{a} = \mathbf{M}^{-1}\,\mathbf{b} \qquad (8)$$

Importantly, whereas the arrays **a** and **b** depend on the specifics of a particular velocity map imaging experiment, the matrices **M** and **M⁻¹** do not, and only depend on the maximum values of $v_{2D/3D}$ (i.e., the number of pixels that the camera image extends from the center corresponding to zero velocity) and the maximum $l_{max}$ that need to be considered. Hence, once these matrices have been determined, they can be used repeatedly to perform inversions.

In the analysis of the experiments reported in this paper, $l_{max}$=6 was used, and the experimental images were inverted within a radius of 110 pixels around their center. The data reported in the paper is based on the Abel inversion of 75.742 images, recorded for different values of $\tau_{XUV-XUV}$ and $\tau_{XUV-NIR}$. $\tau_{XUV-XUV}$ was varied between 11 and 102 fs (in steps of 3 fs), and $\tau_{XUV-NIR}$ was varied between -50 and +800 fs in steps of 4 fs. For each $\tau_{XUV-XUV}$ and $\tau_{XUV-NIR}$, up to 8 pump-probe scans were performed, where data was collected for 2000 laser shots at each delay. The error bars shown in Figure 3 of the main manuscript are based on a statistical analysis of the FTPS of all individual scans, moreover distinguishing the data recorded on the right side and the left side of the image symmetry axis.

**Density matrix considerations for the occurrence of ion-photoelectron entanglement**

The XUV spectral amplitude $F_{XUV}(\omega_{XUV})$ depends on the time delay between the two XUV replicas, and is given by

$$F_{XUV}(\omega_{XUV}) = 2F_{0,XUV}(\omega_{XUV})\cos\left(\frac{\omega_{XUV}\tau_{XUV-XUV}}{2}\right) \qquad (9)$$

where $F_{0,XUV}(\omega_{XUV})$ is the spectral amplitude of each individual XUV pulse, $\omega_{XUV}$ is the XUV frequency and $\tau_{XUV-XUV}$ is the two-pulse delay. As a result, the XUV spectral intensity ($\sim |F_{XUV}(\omega_{XUV})|^2$) contains fringes with a spacing $\Delta\omega = \frac{2\pi}{\tau_{XUV-XUV}}$.

The ion + photoelectron state that is formed in the experiment consists of combinations of ionic vibrational states $v$ and photoelectrons that for simplicity are assumed to be fully characterized by their kinetic energy $\varepsilon$, subject to the condition that the sum of the ionic energy and the electron kinetic energy as provided by the XUV photon, must be contained within the bandwidth of the XUV laser. Within this approximation, the wave function can be written as

$$\Psi_{ion+photoelectron} \sim \sum_v \int d\varepsilon \; \psi_{ion,v}\psi_{photoelectron,\varepsilon} F_{0,XUV}(\varepsilon + E_v)\cos\left[\frac{(\varepsilon + E_v)\tau_{XUV-XUV}}{2}\right] \doteq$$
$$\doteq \sum_v \int d\varepsilon \; \psi_{v\varepsilon}$$
$$(10)$$

where $\psi_{\text{ion},v}$ is the wave function of an $H_2^+$ cation in vibrational state $v$, which lies at an energy $E_v$ with respect to the neutral ground state, and $\psi_{\text{photoelectron},\varepsilon}$ is the wave function of a photoelectron with kinetic energy $\varepsilon$.

The ionic reduced density matrix elements $\rho_{vv\prime}$ are given by

$$\rho_{vv\prime} = \int d\varepsilon\, \psi_{v\varepsilon} \psi_{v\prime\varepsilon}^* \sim \int d\varepsilon\, F_{0,\text{XUV}}(\varepsilon + E_v) \cos\left[\frac{(\varepsilon + E_v)\tau_{\text{XUV-XUV}}}{2}\right]$$
$$F_{0,\text{XUV}}^*(\varepsilon + E_{v\prime}) \cos\left[\frac{(\varepsilon + E_{v\prime})\tau_{\text{XUV-XUV}}}{2}\right]$$
$$= \frac{1}{2} \int d\varepsilon\, F_{0,\text{XUV}}(\varepsilon + E_v)\, F_{0,\text{XUV}}^*(\varepsilon + E_{v\prime})$$
$$\cos\left[\frac{\Delta E_{vv\prime}\tau_{\text{XUV-XUV}}}{2}\right] \cos\left[\left(\frac{(E_v + E_{v\prime})}{2} + \varepsilon\right)\tau_{\text{XUV-XUV}}\right]$$

(11)

where $\Delta E_{v,v\prime}$ is the energy separation between levels $H_2^+(v)$ and $H_2^+(v\prime)$. From this it follows that the vibrational coherence described by $\rho_{vv\prime}$ is zero whenever

$$\frac{\Delta E_{v,v\prime}\, \tau_{\text{XUV-XUV}}}{2} = \frac{(2n+1)\pi}{2} \qquad (12)$$

which agrees with the condition for entanglement given in the main manuscript.